\documentclass[aps,prb,twocolumn,floatfix,showpacs]{revtex4}
\usepackage{amsmath}
\usepackage{graphics}

\newcommand{\vect}[1]{\boldsymbol{#1}}
\newcommand{\aop}{\hat{a}}
\newcommand{\figwidth}{7.75cm}
\newcommand{\Figwidth}{16.4cm}

\begin{document}

\title{Coherent rotations of a single spin-based qubit in a single
  quantum dot at fixed {Z}eeman energy}

\author{Jordan Kyriakidis}
\homepage{http://soliton.phys.dal.ca}
\author{Stephen J. Penney}
\affiliation{Department of Physics and Atmospheric Science, Dalhousie
  University, Halifax, Nova Scotia, Canada, B3H 3J5}

\date{\today}

\begin{abstract}
  Coherent rotations of single spin-based qubits may be accomplished
  electrically at fixed Zeeman energy with a qubit defined solely
  within a single electrostatically-defined quantum dot; the
  $g$-factor and the external magnetic field are kept constant.  All
  that is required to be varied are the voltages on metallic gates
  which effectively change the shape of the elliptic quantum dot.  The
  pseudospin-1/2 qubit is constructed from the two-dimensional
  $S=1/2$, $S_z=-1/2$ subspace of three interacting electrons in a
  two-dimensional potential well.  Rotations are created by altering
  the direction of the pseudomagnetic field through changes in the
  shape of the confinement potential.  By deriving an exact analytic
  solution to the long-range Coulomb interaction matrix elements, we
  calculate explicitly the range of magnitudes and directions the
  pseudomagnetic field can take.  Numerical estimates are given for
  {GaAs}.
\end{abstract}


\maketitle

\section{Introduction}
\label{sec:introduction}

The issue of real-time coherent control of individual quantum states
is central to quantum computing and to many other ideas in the
burgeoning field of quantum nanoelectronics.  The main challenge is to
isolate a system (or the interesting parts of a system) from its
environment in order to prevent decoherence, yet have it
environmentally coupled enough in order to perform measurements to
determine what state (or distribution of states) the system is in.  In
a solid state system---semiconductors in particular---encoding
information in the spin, rather than the charge, of an electron is a
promising path since spin couples more weakly to the environment than
does charge.  But precisely because of this weaker environmental
coupling, controlling (and measuring) the dynamics through external
fields is slower and more problematic than in charge systems.

For quantum computing, employing the spin as the basic qubit, for
essentially the reasons mentioned above, was recognized early
on.\cite{loss98:quant.comput.quant.dots}  Here, the two-qubit gates are
controlled electrically,\cite{burkar99:coupl.quant.dots} but single
qubit rotations---a necessary ingredient in universal quantum
computing---require local fields (or, more precisely, local Zeeman
tuning), and necessitates breaking the spin symmetry explicitly.  In
contrast, one can define \emph{coded}
qubits;\cite{bacon00:univer.fault.toler,kempe01:theor.decoh.free}
rather than defining a logical qubit as being a single electron (or
excess electron) in a single quantum dot, a single logical qubit may
be defined, for example, as several quantum dots.  Explicit gate
sequences~\cite{divin00:univer.quant.comput} for three electrons
respectively confined to three quantum dots explicitly show that the
exchange interaction, controlled through gates (\textit{i.e.},
electrical means) alone is sufficient.  This requires both additional
gates and an order-of magnitude increase in gate operations.

In the present paper, we show how a spin-based qubit, defined in a
\emph{single} quantum dot, may be manipulated exclusively by pulsing
voltages applied to gates; the external magnetic field and the
$g$-factor are uniform, isotropic, and static.  Thus, both single- and
double-qubit gates can be constructed solely through voltage pulsing
with a homogeneous, static Zeeman energy.

\section{Summary}
\label{sec:summary}

Our qubit is encoded in the two-dimensional $S=1/2$, $S_z=-1/2$
subspace of three interacting electrons in a two-dimensional potential
well.  Rotations are created by tuning the eccentricity of the
elliptic confinement potential.

Any two-level system can be described as a pseudospin-1/2
object in a pseudomagnetic field with a Hamiltonian written as
\begin{equation}
  \label{eq:12}
  \hat{H}_{\text{qubit}} = b_x\hat{\sigma}_x +
  b_y\hat{\sigma}_y + b_z\hat{\sigma}_z.
\end{equation}
(The most general Hamiltonian will have an additional term
proportional to the identity operator.)  The $\hat{\sigma}_i$ are the
Pauli spin matrices, and the $b_i$ are parameters dependent upon the
details of the problem.  To rotate qubits, at least one of the three
pseudo\-field components must be tunable; in principle, this degree of
control can be arbitrarily small.
As shown below, the pseudo\-field for the present system lies in a
plane, which we take to be the $x$-$z$ plane ($b_y=0$).  In
particular, we consider pseudo\-field switching between two values,
$\vect b_0$ and $\vect b_1$, which differ in magnitude and in
direction $\theta$.

The crucial point demonstrated below is that the Hamiltonian of
Eq.~\eqref{eq:12} may be realized in a single elliptic quantum dot,
where $b_x$, $b_y$, and $b_z$ all have a \emph{different functional
  dependence} on the eccentricity of the quantum dot.\cite{isoqubit}
Since this eccentricity is tunable by external
gates,\cite{kyriak02:voltag.tunab.singl} the spin-based qubit may be
rotated solely through external gate potentials which are local to the
quantum dot.

Although our results below are for two-dimensional elliptic
confinement, the general scheme holds equally well for \emph{any}
anisotropic (non-circular) confinement potential.  The general
requirements are guided by three considerations.  First, the two qubit
states $|0\rangle$ and $|1\rangle$ should both have the same spin
($1/2$) and spin projection.  Second, if the two states
differ by at least one spin-flipped pair, the relaxation should then
be governed by the spin (rather than charge) relaxation time,
regardless of the orbital configurations.  Third, if those spin-1/2
states which define the qubit are the two lowest-energy states, then
one can serve as the initial state, prepared by equilibration.

In the following section, we outline an exact solution to the one-body
problem.  This solution has been published
before,\cite{madhav94:elect.proper.anisot} but we provide an alternate
derivation based on Bose operators, similar to the circular case,
which will facilitate the second-quantized treatment with
interactions.

In section~\ref{sec:coul-matr-elem}, we consider interactions.  We
provide, for the first time, an exact, closed-form expression for all
Coulomb matrix elements (in the single-particle eigenbasis), valid for
arbitrary quantum numbers.

We next detail the explicit construction of our qubit in
section~\ref{sec:qubit-construction}, and derive Eq.~\eqref{eq:12},
giving expressions for the pseudofields in terms of the various
exchange energies, and, ultimately, in terms of the parameters
appearing in the electronic Hamiltonian.

Following this, we give an explicit sequence of confinement
deformations which enables a qubit flip and give estimates based on
GaAs lateral dots using realistic potential and material parameters.

\section{One-body Hamiltonian: Exact solution}
\label{sec:hamiltonian}

The Hamiltonian for a noninteracting elliptic quantum dot is given by
\begin{equation}
  \label{eq01}
  \hat{H} = \frac{1}{2m} \left( \hat{\vect{p}} - \frac{e}{c}
    \hat{\vect{A}} \right)^2 +
  \frac{1}{2} m \left( \omega_x^2 \hat x^2 + \omega_y^2 \hat y^2 \right).
\end{equation}
We have neglected the Zeeman term since it plays no significant role
in what follows.  Equation~\eqref{eq01} describes one electron trapped
in a plane, under a perpendicular magnetic field---we use the
symmetric gauge, $\hat{\vect{A}} \equiv B (-\hat y,\hat x,0) /
2$---with further lateral confinement by \emph{two different}
parabolic potentials with frequencies $\omega_{x}$ and $\omega_{y}$.
This describes an elliptic confinement with the rotational symmetry
(and consequent angular-momentum conservation) explicitly broken.

Equation~\eqref{eq01} may be diagonalized by introducing Bose
operators analogous to the isotropic case.  (For an alternative but
equivalent solution to the elliptic one-body problem, see
Ref.~\onlinecite{madhav94:elect.proper.anisot}).  These operators are
explicitly given by
\begin{subequations}
  \label{eq:a_ops}
  \begin{gather}
    \begin{pmatrix}
      \aop_1 \\ \aop_2^\dagger
    \end{pmatrix}
    = \frac{1}{\sqrt{2}} 
    \left[ \mathsf{X}\mathsf{Y}^T
      \begin{pmatrix}
        \hat{x}/2\ell_0 \\ \hat{p}_y\ell_0/\hbar
      \end{pmatrix}
      + i \mathsf{X}^{-1}\mathsf{Y}
      \begin{pmatrix}
        \hat{p}_x\ell_0/\hbar \\ \hat{y}/2\ell_0
      \end{pmatrix}
    \right],
    \\
    \mathsf{X} = 
    \begin{pmatrix}
      \alpha_+ & 0 \\ 0 & 1/\alpha_-
    \end{pmatrix},
    \quad\quad
    \mathsf{Y} = 
    \begin{pmatrix}
      \beta_+ & \beta_- \\
      -\beta_- & \beta_+
    \end{pmatrix},
  \end{gather}
\end{subequations}
from which the adjoint operators $(\aop_1^\dagger,\ \aop_2)$ can
easily be found.
These four operators satisfy the canonical Boson commutation
relations.  The dimensionless parameters $\alpha_\pm$, $\beta_\pm$ are
defined by
\begin{subequations}
  \label{eq:alpha-beta}
  \begin{gather}
    \label{eq:4}
    \alpha_\pm = \left( \frac{\omega_0^2 \pm \left( \Omega^2 +
          \omega_-^2 \right)} {\omega_0^2 \pm \left( \Omega^2 -
          \omega_-^2 \right)}
    \right)^{1/4},\\
    \label{eq:9}
    \beta_\pm = \left( 1 \pm \frac{\omega_-^2}{\Omega^2} \right)^{1/2},
  \end{gather}
\end{subequations}
and we have also defined the (hybrid) magnetic length $\ell_{0}^{2} =
\hbar/(m\omega_0)$, cyclotron frequency $\omega_{c} = eB/(mc)$, as
well as~\cite{wxwy}
\begin{gather}
  \label{eq:5}
  \omega_0 = \left[ \omega_c^2 + 2 \left( \omega_x^2 + \omega_y^2
    \right) \right]^{1/2}, \quad
  \omega_- = \left(\omega_x^2 - \omega_y^2\right)^{1/2},\\
  \Omega = \left(\omega_-^4 + \omega_c^2\omega_0^2\right)^{1/4}.
\end{gather}
The Bose operators of Eq.~\eqref{eq:a_ops} diagonalize the elliptic
Hamiltonian, Eq.~\eqref{eq01}:
\begin{equation}
  \label{eq08}
  \hat{H} = \hbar\Omega_+ \left( \aop_1^\dag\aop_1 + \frac{1}{2}
  \right) + \hbar\Omega_- \left( \aop_2^\dag\aop_2 + \frac{1}{2}
  \right),
\end{equation}
where $\Omega_\pm = \frac{1}{2} \sqrt{\omega_0^2 + \omega_c^2 \pm
  2\Omega^2}$.  (In the isotropic limit of $\omega_x \rightarrow
\omega_y$, we have $\alpha_\pm \longrightarrow 1$ and $\beta_\pm
\longrightarrow 1$. The Bose operators and the Hamiltonian then reduce
to the usual isotropic ones.\cite{jacak97:quant.dots})

\section{Coulomb matrix elements: Exact solution}
\label{sec:coul-matr-elem}

For the electron interactions, we use the long-range Coulomb energy
($\sim 1/r$) and work in the second quantized formalism using the
exact single-particle basis $|mn\rangle$ ($\aop_1^\dag\aop_1
|mn\rangle = n |mn\rangle$, $\aop_2^\dag\aop_2 |mn\rangle = m
|mn\rangle$); hence $\hat{V}_C = \frac{1}{2} \sum V_{ijkl}
c^\dag_{i\sigma} c^\dag_{j\sigma'} c_{l\sigma'} c_{k\sigma}$, where
all indices ($ijkl\sigma\sigma'$) are summed over; each Latin index
represents a \emph{pair} of orbital quantum numbers ($m,n$) and the
Greek indices represent spin ($\sigma, \sigma' = \pm 1/2$).
Calculation of the matrix element $V_{ijkl}$ proceeds through the
two-dimensional Fourier transform,~\cite{states}
\begin{equation}
  \label{eq:7}
  V_{ijkl} = \int\!d^2q\, \frac{e^2}{2\pi q}
  (m_1n_1,m_2n_2| 
  \text{e}^{i\vect{q}\cdot(\hat{\vect{r}}_1 - \hat{\vect{r}}_2)}
  |m_3n_3,m_4n_4),
\end{equation}
by writing the position operator
$\hat{\vect{r}} = (\hat x, \hat y)$ in terms of the Bose operators in
Eq.~\eqref{eq:a_ops} and their adjoint.  After some calculation, we
obtain
\begin{widetext}
  \begin{multline}
    \label{eq:v-coul}
    V_{ijkl} =
    \frac{e^2 / (2 \pi \ell_0)}{\sqrt{2 \prod_{k=1}^4 m_k!\, n_k!}}
    \sum_{p_1=0}^{\text{min}(n_1,n_3)} p_1!
    \begin{pmatrix} n_1\\p_1 \end{pmatrix}
    \begin{pmatrix} n_3\\p_1 \end{pmatrix}
    \sum_{p_2=0}^{\text{min}(m_1,m_3)} p_2!
    \begin{pmatrix} m_1\\p_2 \end{pmatrix}
    \begin{pmatrix} m_3\\p_2 \end{pmatrix} \\ \mbox{} \times 
    \sum_{p_3=0}^{\text{min}(n_2,n_4)} p_3!
    \begin{pmatrix} n_2\\p_3 \end{pmatrix}
    \begin{pmatrix} n_4\\p_3 \end{pmatrix}
    \sum_{p_4=0}^{\text{min}(m_2,m_4)} p_4!
    \begin{pmatrix} m_2\\p_4 \end{pmatrix}
    \begin{pmatrix} m_4\\p_4 \end{pmatrix}
    (-1)^{q^-} \Gamma \left(q^+ + {\textstyle \frac{1}{2}}\right) 
    \int_{-1}^1 \! dx \, \frac{\lambda + \lambda^*}{\sqrt{1 - x^2}},
  \end{multline}
\end{widetext}
where $q^{\pm} = \frac{1}{2} \sum_{i=1}^4 (\pm1)^{i-1} (m_i + n_i -
2p_i)$.  The integral may be expressed as a sum of elementary
functions and complete elliptic integrals of the first, second, and
third kinds.  The function $\lambda = \lambda(x)$ is explicitly given
by
\begin{equation}
  \label{eq:lambda}
  \lambda = \frac{u^{n_{12}} (u^*)^{n_{34}} v^{m_{34}}
    (v^*)^{m_{12}}}{\left(|u|^2 + |v|^2\right)^{q^+ + \frac{1}{2}}},
\end{equation}
where $n_{ij} = n_i + n_j - (p_1 + p_3)$, $m_{ij} = m_i + m_j - (p_2 +
p_4)$, and
\begin{equation}
  \begin{pmatrix}
    u \\ v
  \end{pmatrix}
  =
  \begin{pmatrix}
    \beta_+/\alpha_+ & i \alpha_+\beta_-\\
    \alpha_- \beta_- & i \beta_+ / \alpha_-
  \end{pmatrix}
  \begin{pmatrix}
    x \\ \sqrt{1 - x^2}
  \end{pmatrix}.
\end{equation}
The matrix element, Eq.~\eqref{eq:v-coul}, vanishes if $\sum_i (m_i +
n_i)$ is odd and is real otherwise.  In the isotropic
limit~\cite{kyriak02:voltag.tunab.singl} ($\omega_x = \omega_y$) the
expression simplifies considerably and conservation of angular
momentum emerges explicitly.  Equation~\eqref{eq:v-coul} is an
\emph{exact} result, valid for any set of quantum numbers $m_i,\ n_i\ 
(i = 1, \ldots, 4)$.  It can be used as the basis of a numerical
treatment of the many-body problem.

\section{Qubit construction}
\label{sec:qubit-construction}

Rotations are enabled through the mutual exchange interactions among
the confined electrons.  In what follows, we consider three-particle
antisymmetric state vectors of the form $|m_1 n_1 \sigma_1, m_2 n_2
\sigma_2, m_3 n_3 \sigma_3\rangle$, with fixed orbital states
$(m_i,n_i)$.  For a given set of orbital quantum numbers, we construct
the qubits from the (exact) two-dimensional subspace of the
three-electron problem with spin $S=1/2,\ S_z=-1/2$.  We shall
consider the three orbital states $(m,n)=(0,0),\ (1,0),\ (2,0)$ with
no double occupancy.  We stress, however, that neither single
occupancy nor three orbital states (only) are essential to the main
conclusions.  The important point is that the spin-degenerate space is
two-dimensional---an exact result---and that the shape of the dot is
tunable---an experimentally demonstrated
fact.\cite{kyriak02:voltag.tunab.singl} The resulting
eight-dimensional Hilbert space is spanned by the antisymmetrised
(Slater determinant) states $|00\sigma_0, 10\sigma_1,
20\sigma_2\rangle$, which we will simply write as $|\sigma_0,
\sigma_1, \sigma_2\rangle$ (but note that these are
\emph{antisymmetrised} states).  Three spin-1/2 particles can be
combined to form a spin-3/2 quartet and two orthogonal spin-1/2
doublets.  The two $|S,S_z\rangle = |1/2, -1/2\rangle$ states are
orthogonal and form our two qubit states $|0\rangle$ and $|1\rangle$.
They are explicitly given by
\begin{subequations}
  \label{eq:21}
  \begin{gather}
    \label{eq:10}
    |0\rangle \equiv \frac{1}{\sqrt{6}} \left(
      2|\downarrow\downarrow\uparrow\rangle -
      |\downarrow\uparrow\downarrow\rangle -
      |\uparrow\downarrow\downarrow\rangle \right),
    \\ \label{eq:11}
    |1\rangle \equiv \frac{1}{\sqrt{2}} \left(
      |\downarrow\uparrow\downarrow\rangle -
      |\uparrow\downarrow\downarrow\rangle \right).
  \end{gather}
\end{subequations}
These states are linear combinations of single-determinant state
vectors and, as such, go beyond the standard Hartree-Fock treatment.
What's more, at finite magnetic field, these states are both lower in
energy than spin-1/2 states involving a doubly occupied $s$-shell.

We project the total Hamiltonian---consisting of both one-body,
Eq.~\eqref{eq08}, and two-body, Eq.~\eqref{eq:v-coul}, terms---down to
our two-dimensional qubit subspace, spanned by the vectors $|0\rangle$
and $|1\rangle$.  This can be mapped to a pseudo\-spin-1/2 problem
whose general form is given by Eq.~\eqref{eq:12}.  The pseudomagnetic
field components are given by various exchange interactions.  We find
$b_y = 0$, whereas
\begin{subequations}
  \label{eq:14}
  \begin{gather}
    \label{eq:13}
    b_x = \frac{\sqrt{3}}{2} \left( V_{0220} - V_{1221} \right),
    \\
    \label{eq:16}
    b_z = -V_{0110} + \frac{1}{2} \left(V_{1221} + V_{0220}\right).
  \end{gather}
\end{subequations}
The pseudo\-fields $b_x$ and $b_z$ depend on different combinations of
exchange-interaction matrix elements, and each of these depends
differently on the ratio $r \equiv \omega_y/\omega_x$.  This will be
true of almost any anisotropic confinement potential.  Because of
this, the direction of the pseudo\-field can be changed---inducing
coherent rotations of the qubit---by changing the anisotropy parameter
$r$.  Analytic expression for the various exchange energies in
Eq.~\eqref{eq:14} are given in the Appendix.

Figure~\ref{fig:angle} shows the angle $\theta$ of the pseudo\-field
$\vect{b}$ (relative to the positive $x$ axis) as a function of both
anisotropy $r$ and (actual) magnetic field $z \equiv \omega_c /
\omega_x$.
\begin{figure*}
  \resizebox{\Figwidth}{!}{\includegraphics*{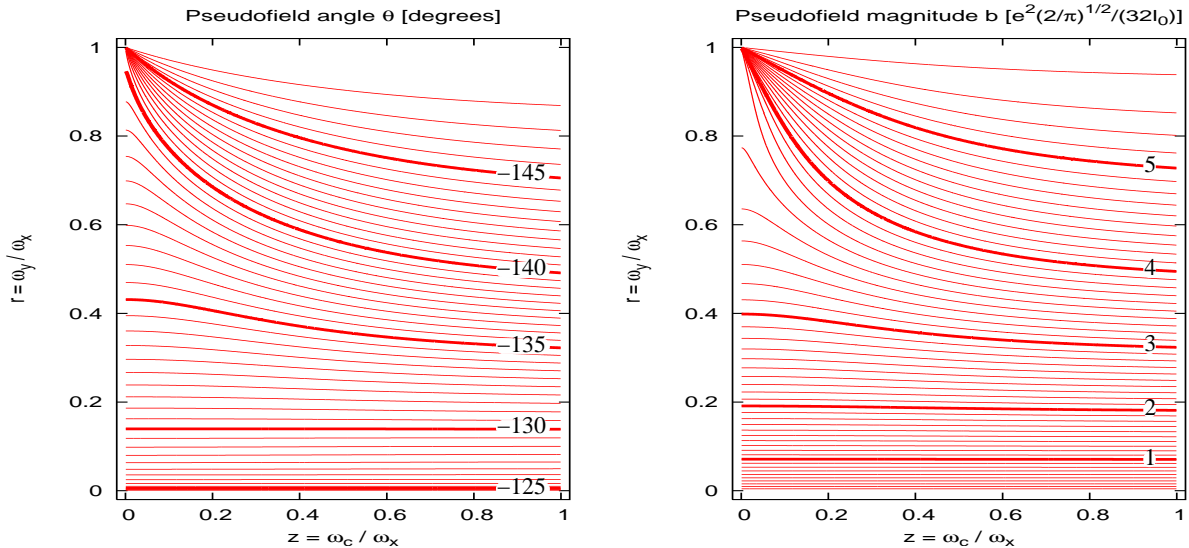}}
  \caption{\label{fig:angle}Contour plot showing the angle $\theta$
    (left plot) and the magnitude $b$ (right plot) of the
    pseudomagnetic field $\vect{b}$ as a function of quantum dot
    anisotropy $r = \omega_y / \omega_x$ and (actual) magnetic field
    $z = \omega_c / \omega_x$.  The angle $\theta$ is measured from
    the positive $x$ axis and $\vect{b}$ lies in the $x$-$z$ plane.
    Angles are measured in degrees and magnitudes in units of
    $e^2\sqrt{2 / \pi} / (32 \ell_0)$.  Shown are lines of constant
    $\theta$ and $b$ respectively.}
\end{figure*}
The larger values of $r$ are the physically relevant ones.  (The
isotropic case, $\omega_x = \omega_y$ corresponds to $r=1$, whereas
$r=0$ is the one-dimensional limit.)  The figure shows that at a
\emph{fixed} magnetic field $z$, a range of pseudo\-field directions
are available for qubit rotations by varying the voltage-tuned
anisotropy $r$.  In both extremes, $r=0,1$, Fig.~\ref{fig:angle} shows
no dependence on $\theta$ with magnetic field $z$; in both
cases, the system essentially has only one tunable parameter which, in
the logical qubit space, tunes the \emph{magnitude} of the
pseudo\-field (through the hybrid magnetic length $\ell_0$).
Figure~\ref{fig:angle} also shows how the magnitude (in units of
$e^2\sqrt{2 / \pi} / (32 \ell_0)$) of the pseudo\-field changes as a
function of $r$ and $z$.  In general, both the magnitude and direction
of the pseudo\-field are altered by the anisotropy.

\section{Explicit qubit flip sequence}
\label{sec:expl-qubit-rotat}

By tuning $r(t)$ in real time, a qubit flip $|0\rangle \rightarrow
|1\rangle$ can be performed; we give here an explicit example.  It is
useful to rotate our qubit, Eq.~\eqref{eq:21}, so that it is oriented
parallel (and antiparallel) to the direction of the pseudo\-field for
$r=1$, given explicitly by
\begin{equation}
  \label{eq:isotrop}
  \vect{b}_0 = \frac{-\sqrt{\pi}}{512} \left( \frac{57}{\sqrt{3}}, 0,
    21 \right) \frac{e^2}{\ell_0}.
\end{equation}
Thus, our rotated qubit states are $|\tilde 0\rangle = c_- |0\rangle -
c_+ |1\rangle$ and $|\tilde 1\rangle = c_+ |0\rangle + c_- |1\rangle$,
where $c_\pm = \sqrt{(b_0 \pm b_{0z}) / (2b_0)}$.

The initial ($t=0$) qubit state is along the pseudo\-field direction
$\vect{b}_0$ given by $r_0=1$, which, in our rotated frame, we take to
lie along the $z$ axis.  The field is then pulsed~\cite{optimal} to a
new value $\vect{b}_1$ given by, $r_1 < 1$.  (This field lies in the
$x$-$z$ plane.)  The qubit will precess about $\vect{b}_1$ with period
$T_1 = \pi\hbar/b_1$.  Half a period later, at $t=T_1/2$, the qubit is
again in the $x$-$z$ plane, whereupon the field is pulsed back to
$\vect{b}_0$.  The qubit precess about this new field with period $T_0
= \pi\hbar / b_0$.  Half a period later, at $t = (T_1 + T_0)/2$, the
field is again pulsed to $\vect{b_1}$ and the process is repeated
every half period.  (Actually, the pseudo\-field does not need to be
switched every half period; an odd number of half-periods suffices.)
If the angle $\mu$ between $\vect{b}_0$ and $\vect{b}_1$ is chosen
such that $\mu = \pi/(2k)$, where $k$ is an integer, the qubit may be
flipped by $k$ pulses at $\vect{b}_1$ with pulse width $T_1/2$, each
separated by an interval $T_0/2$ at $\vect{b}_0$.  The total switching
time is $t^{\text{flip}}_k = kT_1/2 + (k-1)T_0/2$ and can be very
fast.  (See below).  The qubit can in fact cover the Bloch sphere by
judicious choice of pseudo\-fields, which are entirely controlled by
the quantum dot anisotropy.

\begin{figure}
  \centering
  \resizebox{\figwidth}{!}{\includegraphics*{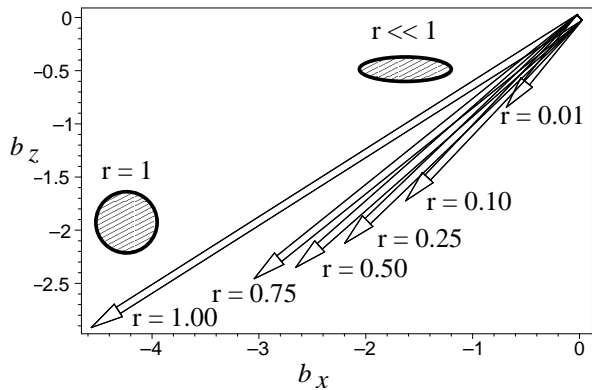}}
  \caption{Direction and magnitude of magnetic field ($b_y = 0$) for
    various anisotropy values $r = \omega_y / \omega_x$ at fixed
    magnetic field $\omega_c = 0.12 \omega_x$.  This is essentially a
    plot of Eq.~\eqref{eq:14} in units of $e^2\sqrt{2 / \pi} / (32
    \ell_0)$.  The hatched circle and ellipse are schematics of the
    quantum-dot shape at different anisotropies.}
  \label{fig:vector}
\end{figure}

For definiteness, we give here numerical estimates based on material
parameters for GaAs.  We take $\omega_x = 6$~meV, while $\omega_y$
switches between 3 and 6~meV.  We also take a (fixed, uniform)
magnetic field of $B = 0.42$~T.  Thus, $r = 1,\ 0.5$ and $z = 0.12$
for GaAs.  For $r=1$, the pseudo\-field is explicitly given by
Eq.~\eqref{eq:isotrop} and yields a magnitude of $b_0 \approx
1.61$~meV.  At $r=0.5$ the magnitude is decreased, $b_1 \approx
0.94$~meV, whereas the direction $\theta$ is increased.
Figure~\ref{fig:vector} shows both the direction and magnitude of the
pseudo\-field for these particular parameters.
The field $\vect{b}_1$ is tilted away from $\vect{b}_0$ by $\mu =
9^\circ$.  This gives a qubit flip in ten pulses.  With these
pseudo\-field values, the precession periods are $T_0 = 1.3$~ps for
$\vect{b}_0$ and $T_1 = 2.2$~ps for $\vect{b}_1$.  The \emph{lower
  bound} on the flipping time $t^{\text{flip}}$ is for a pseudo\-field
switch every half-period; this yields $t^{\text{flip}} \approx
16.8$~ps.  These times are closer to optical frequencies than what is
currently achievable using pulse generators.  Recent pulsed-gate
experiments~\cite{fujis03:elect.pulse.measur} employed electrical
pulse-widths on the order of 10~ns.  With such pulse generators, we
have $t^{\text{flip}} \approx 190$~ns.

\section{Discussion}
\label{sec:discussion}

Our qubit, Eq.~\eqref{eq:21}, has been constructed from a linear
combination of single-determinant (Hartree-Fock) state vectors, where
the orbital degrees of freedom have been frozen out.  But the general
scheme is certainly \emph{not} limited to our specific state vectors.
In general, each logical qubit state can be written as a correlated
many-body state $|Q\rangle = \sum_i \alpha^{(Q)}_i |\psi\rangle_i$,
where $|Q\rangle = |0\rangle, |1\rangle$ is the logical qubit state
and the $|\psi\rangle$ are antisymmetrised orthonormal states,
$|\psi\rangle = |m_1n_1\sigma_1, m_2n_2\sigma_2,
m_3n_3\sigma_3\rangle$, such that $|Q\rangle$ is a spin eigenstate
with $S = 1/2,\ S_z = -1/2$.  Equation~\eqref{eq:21}, for example, has
$(m_1n_1,m_2n_2,m_3n_3) = (00,10,20)$ for both $|Q\rangle =
|0\rangle$, and $|Q\rangle= |1\rangle$; the differences between the
two logical states are, in this case, solely due to spin flips and
phase factors of $\pm 1$.  Although there is no \emph{requirement}
that the orbital degrees of freedom are identical for each qubit
state, it is nevertheless advantageous to have the orbital quantum
numbers identical since this will reduce the electromagnetic
fluctuations which would be present if the qubit rotation involved
orbital transitions as well as spin transitions.

It is always possible to define the logical qubit states in such a way
that they differ only by spin flips and relative phases and not by
their orbital quantum numbers.  This statement is not restricted to
the simple (yet relevant) case of that described by Eq.~\eqref{eq:21}.
It is an exact result, valid even for correlated states involving many
Slater determinants.  Thus, voltage fluctuations due to orbital
transitions can be mitigated.

It is also possible to choose the qubit states such that one is the
ground spin-1/2 state and, consequently, state preparation can be a
matter of equilibration.

Finally, the two qubit states will not be energetically degenerate.
Thus, each qubit state will have different transport characteristics;
the magnitude of current through the dot will depend differently on
gate and bias voltages for each of the qubit states.  This may be
exploited to be used as a detection scheme for final readout.

\begin{acknowledgments}
  Acknowledgment is made of fruitful discussions with Marek
  Korkusinski, Daniel Lidar, and especially Guido Burkard.  This work
  was financially supported by NSERC of Canada.
\end{acknowledgments}

\appendix*

\section{Exchange energies}
\label{sec:exchange-energies}

The pseudofield, Eq.~\eqref{eq:14}, is determined by various exchange
energies.  These $V_{ijkl}$ are in turn determined from the exact
expression of Eq.~\eqref{eq:v-coul} with the subscripts $(i,j,k,l) =
(m_1,m_2,m_3,m_4)$ and all $n_i=0$.  For the cases on interest here,
the relevant $V_{ijkl}$ are given by:
\begin{subequations}
  \label{eq:exchange}
  \begin{gather}
    V_{0110} = C X_2, \quad\quad
    V_{0220} = \frac{1}{4}CX_4,\\
    V_{1221} = \frac{1}{2}C\left(\frac{1}{4}X_6 - 2X_4 + 4X_2\right),
  \end{gather}
\end{subequations}
where $C = e^2 / (4 \pi \ell_0)$ is the Coulomb energy scale, $X_s =
2^{(s+3)/2} \Gamma\bigl((s+1)/2\bigr) I_s$, and
\begin{equation}
  \label{eq:Is}
  I_s = \int_0^1 \! du \, \frac{(cu^2 + d)^{s/2}}
  {(1-u^2)^{1/2} (au^2 + b)^{(s+1)/2}}.
\end{equation}
Each $I_s$ is a linear combination of complete elliptic integrals of
the first and second kind\cite{elliptic}
\begin{equation}
  \label{eq:17}
  I_s = A_s K(m) + B_s E(m),
\end{equation}
where $m =(\alpha_+^2 - \alpha_-^2)/\alpha_+^2$, and the coefficients
$A_s$ and $B_s$ are given by
\begin{subequations}
  \begin{align}
    \label{eq:18}
    A_2 &= \frac{c}{a\sqrt{b}},\\
    A_4 &= \frac{1}{a^2\sqrt{b}} \left(c^2 -
      \frac{\nu^2}{3b(a+b)}\right),\\
    A_6 &= \frac{1}{a^3\sqrt{b}}\left(c^3 - \frac{c\nu^2}{b(a+b)} -
      \frac{4\nu^3(a+2b)}{15b^2(a+b)^2}\right),\\
    B_2 &= \frac{\nu}{a\sqrt{b}(a+b)},\\
    B_4 &= \frac{2\nu}{a^2\sqrt{b}(a+b)} \left(c +
      \frac{\nu(a+2b)}{3b(a+b)}\right),\\
    B_6 &= \frac{1}{a^3\sqrt{b}} \left(\frac{3c^2\nu}{a+b} + 
      \frac{2c\nu^2(a+2b)}{b(a+b)^2}
    \right. \notag  \\  
      & \left. \quad\quad\quad\quad\quad\quad
      + \frac{8\nu^3(a+2b)^2}{15b^2(a+b)^3} - 
      \frac{3\nu^3}{5b(a+b)^2}\right),
  \end{align}
\end{subequations}
where
\begin{subequations}
  \begin{align}
    \label{eq:19}
    a &= \beta^2_ + \left(\frac{1}{\alpha^2_+} -
      \frac{1}{\alpha^2_-}\right) + 
    \beta^2_-(\alpha^2_- - \alpha^2_+),\\
    b &= \beta^2_+/\alpha^2_- + \alpha^2_+\beta^2_-,\\
    c &= \alpha^2_-\beta^2_- - \beta^2_+/\alpha^2_-,\\
    d &= \beta^2_+/\alpha^2_-,\\
    \nu &= ad-bc,
  \end{align}
\end{subequations}
and the $\alpha_\pm$ and $\beta_\pm$ are given in
Eq.~\eqref{eq:alpha-beta}.


\begin{thebibliography}{14}
\bibitem[{iso()}]{isoqubit}
\bibinfo{note}{In the istropic limit of a circular quantum dot, we find $b_y=0$
  and $b_x \propto b_z$; only the magnitude, not the direction, of the
  pseudofield can be changed; the direction of the pseudofield can only be
  changed through control of the anisotropy.}
\bibitem[{wxw()}]{wxwy}
\bibinfo{note}{Without loss of generality, we restrict $\omega_x \ge \omega_y$
  in what follows.}
\bibitem[{opt()}]{optimal}
\bibinfo{note}{A smoother pulse shape for $r(t)$, while more realistic, would
  entail a careful treatment of time-ordering and would unnecessarily obfuscate
  the present simple example.}
\bibitem[{ell()}]{elliptic}
\bibinfo{note}{We define elliptic integrals of the first and second kind as \[
  K(m) = \int_0^1 \! dt \, \frac{1}{\sqrt{(1-t^2)(1-mt^2)}} \] and \[ E(m) =
  \int_0^1 \! dt \, \frac{\sqrt{1-mt^2}}{\sqrt{1-t^2}}\] respectively.}
\end{thebibliography}

\end{document}